\RequirePackage[T1]{fontenc}
\documentclass[conference,a4paper]{IEEEtran}
\usepackage{amsmath,amssymb}
\usepackage{booktabs}
\usepackage{flushend}
\usepackage{cite}
\usepackage{url}
\usepackage{tikz}
\usetikzlibrary{arrows.meta}
\newcommand{\R}{\mathbb{R}}
\newcommand{\pos}[1]{[#1]_+}

\begin{document}
\title{Spatial Price Formation in Transport Networks:\\
A Variational Approach}
\author{\IEEEauthorblockN{William F. A. Weber and Mahendra Piraveenan}
\IEEEauthorblockA{School of Computer Science, Faculty of Engineering\\
The University of Sydney, Sydney, Australia}}
\maketitle
\raggedbottom

\begin{abstract}
This paper formulates a model of spatial price formation in a closed regional
economy, with finitely many production nodes and finitely many demand nodes
embedded in a continuous geographic domain. Transportation costs depend on the
geometry of the region and on an underlying network of feasible connections.
Prices are modeled as a spatial field constrained by arbitrage and transport
feasibility. The equilibrium price system is characterized by a variational
problem. A second minimization selects an equilibrium-compatible price field
with minimum spatial energy and permits the use of Euler--Lagrange methods.
The framework clarifies what observed prices and quantities reveal about
demand, and uses known transport costs to bound unobserved prices and exclude
directed arcs from equilibrium shipments. A five-node example illustrates equilibrium,
alternative shipment patterns, and the information supplied by partial price
observations. We give a numerical procedure for planar domains and outline an extension
to spherical domains.
\end{abstract}


\section{Introduction}

Complex networks provide a powerful framework for representing and analysing
systems in which the interactions among individual components give rise to
collective behaviour. Such systems include social networks, communication
and transportation infrastructure, and economic networks, where connectivity,
network topology, and the distribution of flows can significantly influence
system performance and resilience~\cite{newman2003structure,boccaletti2006complex,piraveenan2012congruity}.
Understanding the relative importance of nodes and links is particularly
relevant when studying the propagation of disruptions through interconnected
systems, as demonstrated by research on percolation centrality and network
robustness~\cite{piraveenan2013percolation,kasthurirathna2013robustness}.
In transportation and supply networks, these structural considerations must
also be accompanied by an understanding of the economic interactions between
production and consumption locations, since transport costs and network
connectivity influence commodity flows, market accessibility, and regional
welfare~\cite{piraveenan2023transportation,fajgelbaum2020optimal}.
This motivates the study of spatial price formation on transportation networks,
where the relationships between network structure, transport costs, and economic
equilibrium can be examined within a common mathematical framework.

We consider a closed regional economy occupying a compact geographic domain.
The region is isolated by natural barriers, so that no trade crosses its
boundary and no demand originates outside the region. Within the region there
are finitely many locations at which a homogeneous good is produced, and
finitely many locations at which the good is demanded. Transportation between
locations is costly and depends on the geometry of the region and on a given
network of feasible routes.

The objective of the paper is threefold. First, to provide a variational
formulation of spatial price determination consistent with arbitrage and
transport costs. Second, to clarify what observed prices and quantities,
together with known transport costs, reveal about demand and feasible shipments.
Third, to outline a concrete computational strategy for planar domains,
with an extension to the surface of a sphere.

The analysis treats price as a field over space. Production and demand occur
at finitely many locations, while arbitrage relates prices throughout the
region. A spatial smoothing criterion selects a field consistent with equilibrium
and permits the use of Euler--Lagrange methods as presented in
Gelfand and Fomin~\cite{gelfand1963calculus}.

The model is developed in two stages. The first determines equilibrium prices
through a price minimization. The second selects a spatially
smooth field while retaining the optimal value of the first problem.
Observed prices and quantities then provide information about demand and
possible commodity movements. Our earlier work examines node importance
through the allocation of heterogeneous travelers across transport
paths~\cite{weber2026}. The framework also provides a basis for
studying how disruptions to production and transportation nodes affect
prices, consumption, and regional welfare.

\subsection{Related literature and contribution}

Enke~\cite{enke1951equilibrium} represented equilibrium among separated
markets by an electrical analogue, and Samuelson~\cite{samuelson1952spatial}
related equilibrium prices and shipments to surplus maximization and linear
programming. Takayama and Judge~\cite{takayama1971spatial} developed
mathematical programming formulations of spatial and temporal price and
allocation models. Beckmann's continuous model of
transportation~\cite{beckmann1952continuous} established an early connection
between spatial commodity movement and variational methods.
Nagurney~\cite{nagurney1999network} develops spatial and network equilibria
through variational inequalities. Florian and Los~\cite{florian1982static}
develop network formulations of static spatial price equilibrium.
Fisk~\cite{fisk1987spatial} gives a price formulation on congested networks
using variational inequalities. The relationship between transport costs
and admissible price differences is also central to optimal
transport~\cite{kantorovich1942translocation,villani2003topics}.

Fajgelbaum and Schaal~\cite{fajgelbaum2020optimal} study infrastructure
investment and commodity transport on a network in spatial equilibrium,
using a dual formulation in prices for computation.
Allen and Arkolakis~\cite{allen2022welfare} evaluate the welfare effects of
infrastructure improvements in a spatial general equilibrium model with
endogenous transport costs and congestion.

Metric extension results~\cite{mcshane1934} allow compatible prices at
observed locations to be extended over the region. They also supply bounds
on unobserved prices. We use these bounds to determine which arcs can be
excluded from equilibrium shipments when observations are incomplete.

Piraveenan and Saripada~\cite{piraveenan2023transportation} introduced
Transportation Centrality, which allocates traffic over all simple paths
using a logit function of path cost. Weber and Piraveenan~\cite{weber2026}
extend this measure to heterogeneous traveler classes with toll-sensitive
path costs. Nagurney and Qiang~\cite{nagurney2008efficiency} assess the
importance of nodes and links through the effect of their removal on a
network performance measure incorporating demand, flows, and costs.
The present model concerns a homogeneous commodity and competitive arbitrage.

The equilibrium price formulation builds on classical spatial price equilibrium
theory. We combine it with a minimum-energy spatial selection criterion and
metric-extension bounds from partial price observations. These bounds give
attainable ranges for prices compatible with arbitrage and identify directed
transport arcs that cannot carry equilibrium shipments.

\section{Geographic and Economic Environment}

\subsection{Geographic domain and nodes}

Let $\Omega\subset\R^2$ be a compact, connected region with a smooth boundary.
It represents a region surrounded by an impenetrable mountain range. There
is no economic activity outside $\Omega$. A later extension replaces the
planar region by the sphere $S^2$.

There are two finite, nonempty sets of economic nodes. Production nodes
$A=\{a_1,\ldots,a_M\}$ have capacities $Q_i>0$. Demand nodes
$B=\{b_1,\ldots,b_N\}$ host populations consuming the good. Their locations
are given by observed geographic coordinates. Other locations may serve as transport
junctions without producing or consuming the good.

\subsection{Transportation network}

The cost $c(x,y)$ is the minimum unit transport cost between locations $x$
and $y$, given physical distance, feasible land and sea routes, and exogenous
infrastructure restrictions. We assume that $c$ is a finite, continuous
metric: it is zero only between identical locations, is symmetric, and
satisfies the triangle inequality. Symmetry and metric structure are modelling
assumptions adopted here; extensions to directed networks with asymmetric
transportation costs are left for future work. Routes have constant unit costs and no
binding shipment capacities. An impassable connection is omitted.

For computation, the feasible connections form a connected undirected graph
$G=(V,E)$ embedded in the region. Its vertices include the economic nodes and
transport junctions. An edge or link is an undirected connection; an arc is
one direction of an edge, and a path or route is a sequence of edges or arcs.
Each edge $\{u,v\}$ has a positive cost $w_{uv}$, measured
in the same monetary units per unit of the good as prices. The network cost
between two vertices is
\begin{equation}\label{eq:distance}
 d_G(u,v)=\min_{\text{paths }u\to v}\sum_{\{r,s\}\text{ on path}}w_{rs}.
\end{equation}
A direct connection may cost more than a route through other vertices.
Both its own cost and the cost of the cheapest available route matter when
interpreting observed prices.

We use two related transport settings: a finite graph with specified transport
costs for the economic calculations, and a continuous path-cost formulation
for the spatial Euler--Lagrange calculation. The numerical examples use graph
constraints, with the triangular mesh supplying interpolation and spatial energy.

\section{Economic Primitives}

\subsection{Demand}

At each demand node $b_j$, consumers have unit demand with heterogeneous
valuations. Let $F_j$ denote the cumulative distribution function of
valuations, and let $\mu_j>0$ be the mass of consumers. If the delivered
price is $p$, demand equals
\begin{equation}\label{eq:demand}
 D_j(p)=\mu_j\bigl(1-F_j(p)\bigr).
\end{equation}
Thus $1-F_j(p)$ is the share of consumers willing to pay the price, and
multiplication by $\mu_j$ gives the quantity purchased. Assume each $F_j$ is
continuous, with valuations between zero and a finite common upper bound
$\bar v$.

Consumer surplus is the total willingness to pay of the buyers, less their
expenditure. It is the area under the demand curve above the price:
\begin{equation}\label{eq:surplus}
 CS_j(p)=\int_p^\infty D_j(t)\,dt.
\end{equation}
Its derivative is $CS_j'(p)=-D_j(p)$. Since demand falls as price rises,
consumer surplus is convex in price.

\subsection{Production}

Each production node has zero marginal cost up to capacity $Q_i$.
Its actual supply $s_i$ satisfies $0\le s_i\le Q_i$. At a positive producer
price the full capacity is supplied; at zero price any supply within the
capacity is consistent with revenue maximization. At a negative price the
producer supplies nothing. The producer's maximum revenue is
$Q_i\pos{p(a_i)}$, where $\pos{z}=\max\{z,0\}$.

\subsection{Prices and arbitrage}

Let $p:\Omega\to\R$ denote the price field. If the price difference between two locations exceeded the transport cost
between them, a trader could profit by buying at the lower price and
selling at the higher price. The no-arbitrage
condition is therefore
\begin{equation}\label{eq:arbitrage}
 |p(x)-p(y)|\le c(x,y)\qquad(x,y\in\Omega).
\end{equation}
Let $\mathcal P$ denote the continuous fields satisfying this condition.
On the graph, it suffices to impose $|p_u-p_v|\le w_{uv}$ on each edge:
adding these inequalities along a path gives the corresponding bound on
its endpoints. Consumers at $b_j$ face the delivered price $p(b_j)$.

\section{Variational Formulation}

\subsection{Market clearing and feasible shipments}

Write $f_{uv}\ge0$ for the shipment from $u$ to $v$ along an arc.
Both directions of each physical edge are available. At every vertex,
incoming shipments and local production must equal outgoing shipments and
local consumption:
\begin{equation}\label{eq:balance}
 \sum_{u:\{u,v\}\in E}f_{uv}+s_v
 =\sum_{u:\{u,v\}\in E}f_{vu}+d_v.
\end{equation}
Here $s_v$ and $d_v$ are the total supply and consumption at vertex $v$,
and are zero where the corresponding activity is absent. Summing over
vertices gives total supply equal to total consumption. Production capacities
may remain partly unused when demand is insufficient.

The quantity formulation maximizes the gross benefit of consumption less
transport costs, subject to these balance equations, production capacities,
and consumption bounds $0\le d_j\le\mu_j$. This is the usual spatial
surplus-maximization problem~\cite{samuelson1952spatial,takayama1971spatial}.

\subsection{The equilibrium price problem}

The corresponding price formulation is
\begin{equation}\label{eq:price}
 \min_{p\in\mathcal P}J(p),\qquad
 J(p)=\sum_j CS_j(p(b_j))+\sum_i Q_i\pos{p(a_i)}.
\end{equation}
The first term is consumer surplus; the second is the producers' maximum
revenue. Together they provide an upper bound on the surplus achievable by
moving and consuming the good. Minimizing this bound gives the equilibrium
price problem.

To see the connection, let $U_j(d_j)$ be the gross willingness to pay for
$d_j$ units at node $b_j$, obtained by serving the highest-valued units
first. At a given price, $CS_j(p)+pd_j\ge U_j(d_j)$, with equality when
$d_j=D_j(p)$. Producer revenue satisfies
$p(a_i)s_i\le Q_i\pos{p(a_i)}$. Finally, market clearing expresses consumer
expenditure less producer revenue as the sum of price differences times
shipments. Arbitrage bounds each such difference by its transport cost.
Combining these observations gives
\begin{equation}\label{eq:welfare-bound}
 \sum_j U_j(d_j)-\sum_{(u,v)}w_{uv}f_{uv}\le J(p),
\end{equation}
where the second sum runs over directed arcs.

Equality holds when consumers purchase their demanded quantities, producers
maximize revenue within capacity, and every positive shipment satisfies
\begin{equation}\label{eq:used-route}
 f_{uv}>0\quad\Longrightarrow\quad p_v-p_u=w_{uv}.
\end{equation}
Together with market clearing, these are the competitive equilibrium
conditions. The classical finite-network allocation problem has matching
quantity and price optima; the bound above also gives a direct way to
check a computed equilibrium. A feasible allocation and price field that attain equality in
\eqref{eq:welfare-bound} certify optimality.

The graph version of \eqref{eq:price} is a convex minimization over finitely
many prices with linear edge constraints. The graph and continuous
formulations have the same feasible prices at the economic nodes, and the
same minimum economic value, when $d_G(z,z')=c(z,z')$ for all production
and demand locations $z,z'$. Compatible node prices admit an extension over
the region~\cite{mcshane1934}. Prices between economic nodes can consequently
be chosen by a spatial criterion while preserving the equilibrium value.

\subsection{Spatial selection}

To select the field between nodes, we introduce the Dirichlet energy
\begin{equation}\label{eq:energy}
 \mathcal E(p)=\frac12\int_\Omega|\nabla p(x)|^2\,dx.
\end{equation}
The gradient $\nabla p$ measures the local rate and direction of price
change. Squaring its magnitude penalizes steep changes, and integration
sums this penalty over the region.

Let $J^*$ be the minimum value of the economic problem
\eqref{eq:price}. The spatial selection problem is
\begin{equation}\label{eq:selection}
 \min_{p\in\mathcal P}\mathcal E(p)
 \quad\text{subject to }J(p)=J^*.
\end{equation}
This gives priority to economic equilibrium, then selects an
equilibrium-compatible field with least spatial variation. The smoothing
criterion can select among otherwise undetermined prices; it leaves the
optimal economic value unchanged. Absolute price levels continue to be
constrained by demand and production.

For the continuous smoothing calculation, assume that transport cost
accumulates along paths within $\Omega$ at a positive continuous charge
$\tau(x)$ per unit distance. The metric $c$ is the cheapest such path cost.
The arbitrage condition then bounds the local slope by
$|\nabla p(x)|\le\tau(x)$ wherever the gradient exists. These fields have
finite energy and well-defined values at the nodes.

Under these assumptions, both minimizations have solutions. A bounded value
of $J$ bounds producer prices from above and demand-node prices from below;
arbitrage then bounds prices throughout the region. Continuity of $c$ gives
a common bound on price changes over short distances, so a minimizing
sequence has a uniformly convergent subsequence whose limit attains $J^*$.
For the second minimization, the same price and slope bounds give a uniformly
convergent subsequence of an energy-minimizing sequence. Its limit remains
an economic minimizer and has energy no greater than the limiting energy,
so the second minimum is attained.

The minimizers of $J$ form a convex set. Two minimum-energy fields therefore
have the same gradient almost everywhere and differ by a constant on
connected $\Omega$. The selected field is unique whenever the economic
optimum fixes at least one node price.

\subsection{Euler--Lagrange calculation}

We consider a small open region $U$ containing no production or demand node,
where the slope remains below its permitted value by a positive margin.
Let $h$ be a smooth change to the field that is zero outside a smaller
region inside $U$. For sufficiently small positive or negative
$\varepsilon$, the field $p+\varepsilon h$ remains feasible and has the same
economic value: none of the node prices has changed. At an energy minimum,
the first-order change must be zero:
\begin{equation}\label{eq:variation}
 \left.\frac{d}{d\varepsilon}\mathcal E(p+\varepsilon h)
 \right|_{\varepsilon=0}
 =\int_U\nabla p\cdot\nabla h\,dx=0.
\end{equation}
Integration by parts rewrites the integral as
$-\int_U(\Delta p)h\,dx$. There is no boundary term because $h$ vanishes
near the boundary of $U$. Since the calculation holds for every such
variation, the local Euler--Lagrange equation is
\begin{equation}\label{eq:euler}
 \Delta p=0\quad\text{in }U.
\end{equation}
Here $\Delta p$ is the sum of the second spatial derivatives. A field
satisfying this equation is harmonic: its value at the center of a small
disk equals its average around the disk's boundary. It describes the smooth
interpolation of prices in regions free of economic nodes and binding
arbitrage constraints.

This is the first-variation method described by Gelfand and
Fomin~\cite[Sec.~35]{gelfand1963calculus}. At economic nodes the conditions
come from demand, production, and shipments. Where arbitrage binds, the
permitted price difference is reached and variations must respect that
constraint.

\section{Inference and Initialization from Data}

\subsection{Observed data and node selection}

For empirical applications, we require data on population centers,
production facilities, and transportation routes, together with prices and
quantities observed at a subset of locations. All locations are mapped into a common coordinate system.
Population counts supply consumer masses or a basis for estimating them;
production data distinguish available capacity from realized output.

For an initial empirical model, we would use relatively high population and production thresholds,
yielding small sets of economic nodes. Lowering the thresholds adds nodes
and refines the spatial resolution. Aggregating omitted production and demand
into retained nodes preserves the regional totals. Route distances and travel
times are converted into monetary costs per unit of the good.

\subsection{Demand primitives}

With known consumer mass $\mu_j$, an observed price $p_j$ and quantity $d_j$
give one value of the valuation distribution:
\begin{equation}\label{eq:demand-identification}
 F_j(p_j)=1-\frac{d_j}{\mu_j}.
\end{equation}
Repeated observations at different prices provide further values. For
example, if half of a known population purchases at a given price, that
price is a median valuation under the continuous demand model.

Estimation of a whole demand curve uses repeated observations, a stable
demand specification, and a treatment of price endogeneity. When data are
sparse, parameters may be pooled across nodes or modeled as functions of
observable covariates. Demand elasticities follow from the estimated demand
curve. The Euler--Lagrange calculation describes spatial interpolation
conditional on these demand primitives.

\subsection{Unobserved prices}

Suppose prices $\tilde p_o$ are observed at a nonempty set of vertices $O$.
They are consistent with arbitrage when every observed price difference
is at most the shortest-path cost between those observations. Each observed
price gives both a lower and an upper bound on an unobserved price at $v$.
Combining all observations gives
\begin{equation}\label{eq:price-bounds}
 \begin{aligned}
 L_v&=\max_{o\in O}\{\tilde p_o-d_G(o,v)\},\\
 U_v&=\min_{o\in O}\{\tilde p_o+d_G(o,v)\},\qquad
 L_v\le p_v\le U_v.
 \end{aligned}
\end{equation}
Each observation gives a lower bound by subtracting the transport cost and
an upper bound by adding it. Equation~\eqref{eq:price-bounds} takes the largest
lower bound and the smallest upper bound.
These are the lowest and highest price fields consistent with arbitrage and
the observations~\cite{mcshane1934}. They provide feasible initial fields and
exact individual price ranges.

For two vertices $u,v$, the individual intervals and the cost between the
vertices must both be respected. Their combined restriction is
\begin{equation}\label{eq:difference-bounds}
 \begin{gathered}
 \max\{-d_G(u,v),L_v-U_u\}\le p_v-p_u,\\
 p_v-p_u\le\min\{d_G(u,v),U_v-L_u\}.
 \end{gathered}
\end{equation}
These bounds are attainable among arbitrage-compatible price fields matching
the observations. To see why, choose a price in each of the two
individual intervals. Each chosen price satisfies the arbitrage bounds relative to every
observation. If the difference between the two chosen prices also respects
the transport cost between the two vertices, the enlarged set of
observations is consistent. It therefore extends to the rest of the network. The possible
differences between the intervals run from $L_v-U_u$ to $U_v-L_u$;
intersecting this range with the transport bound gives
\eqref{eq:difference-bounds}, including its endpoints.

\subsection{What prices reveal about arcs}

For an arc $u\to v$, define its slack as
$r_{uv}=w_{uv}-(p_v-p_u)$. A positive slack means that the price difference
falls short of the transport cost. Such an arc carries no positive
equilibrium shipment. At zero slack, the price difference exactly covers
transport cost; the arc may be used, subject to market clearing.
Actual use can still be zero when alternative routes serve the same demand.

Subtracting the price-difference interval in
\eqref{eq:difference-bounds} from $w_{uv}$ gives the exact slack interval.
Arcs are classified across compatible price fields as follows:
\begin{itemize}
 \item Positive minimum slack: excluded from equilibrium shipments.
 \item Zero maximum slack: necessarily tight.
 \item Zero minimum and positive maximum slack: unresolved.
\end{itemize}
These classifications use the observed prices and known transport costs.
An undirected link is excluded only when both of its arcs are excluded.
Known demand and production primitives, together with the equilibrium
conditions, can further restrict the set of admissible prices. For a
fixed, complete equilibrium price field and observed supplies and deliveries,
we obtain shipment ranges by minimizing and maximizing an arc's flow subject
to market clearing and zero flow on slack arcs. Adding consistent price
observations can only narrow the ranges
of feasible prices and arc slack.

Observed prices may also be supplied as intervals reflecting measurement
uncertainty. Minimizing and maximizing each arc's slack under these
intervals and the edge constraints are linear programs. This gives a direct
extension of the same calculation to uncertain observations.

\section{Implementation}

\subsection{Planar implementation}

We construct a triangular mesh over the region and include the economic nodes
as vertices. Price is represented by its values at mesh vertices and is
linearly interpolated across each triangle. A separate graph on these
vertices records feasible transport connections and their costs. The mesh
describes field geometry; the graph describes commodity movement.

The energy of the interpolated field is computed by adding the contribution
of each triangle:
\begin{equation}\label{eq:discrete-energy}
 \mathcal E_h(p)=\frac12\sum_T |T|\,|\nabla p|_T^2.
\end{equation}
Here $|T|$ is triangle area and $\nabla p|_T$ is the constant slope on that
triangle. The subscript $h$ indicates the mesh representation. This is a
quadratic function of the vertex prices, with coefficients determined by
triangle geometry. The construction follows the approximation
method described by Gelfand and Fomin~\cite[Sec.~40]{gelfand1963calculus}.

We use the following computational procedure:
\begin{enumerate}
 \item Specify node locations, producer capacities, consumer masses and
       valuation distributions, the mesh, and edge costs.
 \item Choose feasible initial prices. A constant field suffices for the
       forward model. With observed prices, use a compatible field from
       \eqref{eq:price-bounds} or minimize energy while holding observations
       fixed.
 \item Minimize the economic price objective \eqref{eq:price} over vertex
       prices, enforcing the graph edge constraints, and record its value $J^*$.
 \item Minimize \eqref{eq:discrete-energy} under the same constraints while
       retaining $J(p)=J^*$. This is the spatial selection step.
 \item Compute demand and feasible shipments, and check producer behavior,
       local balance, arbitrage, and equality in \eqref{eq:welfare-bound}.
       Report node prices, quantities, and binding arcs.
\end{enumerate}
To hold observed prices fixed, impose them as constraints in the economic
minimization. They are compatible with equilibrium under the specified
primitives if the constrained minimum equals $J^*$. When the observations
are compatible, fixing at least one observed price makes the minimum-energy
completion unique. The spatial selection step preserves the economic optimum
within the numerical tolerance used for this comparison.

With uniform valuation distributions, equilibrium delivers a unique
quantity at each consumer node. The implementation uses these quantities
to express the optimal-price set with linear constraints; the spatial
step then minimizes a quadratic energy. General demand specifications
use the corresponding constrained convex problems.

A direct discretization of the continuous path-cost model also enforces
the local slope bound on each triangle.

\subsection{Spherical extension}

For the spherical extension, we replace the planar region with the sphere
$S^2$ and use a spherical mesh.
Transport costs come from geodesic distance or feasible network routes.
The energy uses surface area and tangential price gradients. In a region
with slack local transport constraints and no economic nodes, the same
first-variation calculation gives the surface version of the harmonic
equation, using the Laplace--Beltrami operator. The sphere has no boundary;
price levels continue to be constrained by the economic problem. Numerical
examples below use planar meshes.

\section{A Worked Regional Economy}\label{sec:example}

\subsection{Nodes, demand, and transport}

We consider two producers $a_1,a_2$, two demand nodes $b_1,b_2$, and a transport
junction $x$. Producer capacities are $3$ and $5$. Valuations at both demand
nodes are uniform between $0$ and $6$, with consumer masses $6$ and $10$.
For prices between $0$ and $6$, demand is therefore $D_1(p)=6-p$ and
$D_2(p)=10-5p/3$. At prices below zero, demand equals the consumer mass; at prices above six,
it is zero.

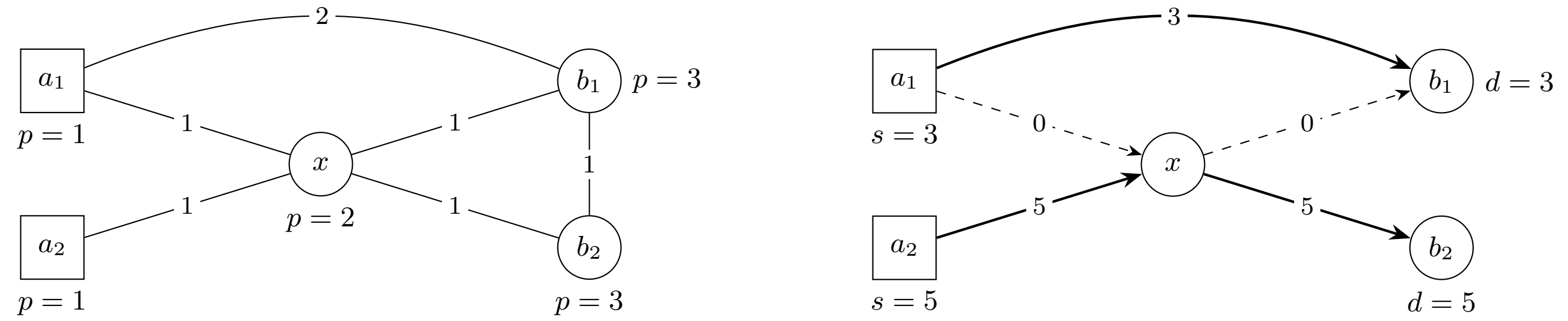
\begin{figure*}[t]
\centering
\begin{minipage}{0.48\textwidth}
\centering
\begin{tikzpicture}[x=1.65cm,y=1.15cm,
 every node/.style={font=\small},
 market/.style={draw,circle,fill=white,minimum size=7mm,inner sep=1pt},
 producer/.style={market,rectangle},
 cost/.style={fill=white,inner sep=2pt,font=\footnotesize}]
 \node[producer,label=below:{$p=1$}] (a1) at (0,1.6) {$a_1$};
 \node[producer,label=below:{$p=1$}] (a2) at (0,0) {$a_2$};
 \node[market,label=below:{$p=2$}] (x) at (1.8,.8) {$x$};
 \node[market,label=right:{$p=3$}] (b1) at (3.6,1.6) {$b_1$};
 \node[market,label=below:{$p=3$}] (b2) at (3.6,0) {$b_2$};
 \draw (a1)--node[cost]{$1$}(x);
 \draw (a2)--node[cost]{$1$}(x);
 \draw (x)--node[cost]{$1$}(b1);
 \draw (x)--node[cost]{$1$}(b2);
 \draw (a1) to[bend left=22] node[cost]{$2$}(b1);
 \draw (b1)--node[cost]{$1$}(b2);
 \node[font=\small] at (1.8,-.85) {(a) Physical graph; edge labels are costs.};
\end{tikzpicture}
\end{minipage}\hfill
\begin{minipage}{0.48\textwidth}
\centering
\begin{tikzpicture}[x=1.65cm,y=1.15cm,>=Stealth,
 every node/.style={font=\small},
 market/.style={draw,circle,fill=white,minimum size=7mm,inner sep=1pt},
 producer/.style={market,rectangle},
 cost/.style={fill=white,inner sep=2pt,font=\footnotesize}]
 \node[producer,label=below:{$s=3$}] (a1) at (0,1.6) {$a_1$};
 \node[producer,label=below:{$s=5$}] (a2) at (0,0) {$a_2$};
 \node[market] (x) at (1.8,.8) {$x$};
 \node[market,label=right:{$d=3$}] (b1) at (3.6,1.6) {$b_1$};
 \node[market,label=below:{$d=5$}] (b2) at (3.6,0) {$b_2$};
 \draw[->,dashed] (a1)--node[cost]{$0$}(x);
 \draw[->,thick] (a2)--node[cost]{$5$}(x);
 \draw[->,dashed] (x)--node[cost]{$0$}(b1);
 \draw[->,thick] (x)--node[cost]{$5$}(b2);
 \draw[->,thick] (a1) to[bend left=22] node[cost]{$3$}(b1);
 \node[font=\small] at (1.8,-.85) {(b) Tight graph; labels show one shipment plan.};
\end{tikzpicture}
\end{minipage}
\caption{The physical network and its equilibrium tight graph. Squares are
producers, $b_1,b_2$ are demand nodes, and $x$ is a junction. All five arcs
in (b) are tight; the two dashed arcs carry zero flow in the displayed plan.
The physical edge $b_1b_2$ is slack in both directions.}
\label{fig:network}
\end{figure*}

The physical network is shown in Fig.~\ref{fig:network}. The four links
joining $x$ to the economic nodes each cost $1$. The direct link $a_1b_1$
costs $2$, and $b_1b_2$ costs $1$. These are all the edges, and each is
available in both directions. Costs are specified independently of the
display coordinates.

\subsection{Equilibrium prices and quantities}

An equilibrium is
\begin{equation}\label{eq:example}
 (p_{a_1},p_{a_2},p_x,p_{b_1},p_{b_2})=(1,1,2,3,3).
\end{equation}
At consumer price $3$, demand is $6-3=3$ at $b_1$ and
$10-5(3)/3=5$ at $b_2$. Both producers receive a positive price and supply
their full capacities. One shipment plan sends three units directly from
$a_1$ to $b_1$, and five units along $a_2\to x\to b_2$.

The plan clears every market. Each unit on the direct arc costs $2$,
equal to the price difference $3-1$. The other five units pass through
two links, each with cost $1$ and price increase $1$. Thus the producer,
consumer, and shipment conditions all hold.

For a further check, gross consumer benefit is $13.5$ at $b_1$ and $22.5$
at $b_2$. These are the areas under the inverse demand curves up to the
delivered quantities. Total transport cost is $3(2)+5(1+1)=16$, so net
surplus is $36-16=20$. Consumer surplus at price $3$ is $4.5+7.5=12$,
and the producer term in \eqref{eq:price} is $3(1)+5(1)=8$.
The price objective is also $20$, establishing equality in
\eqref{eq:welfare-bound}.

\subsection{Alternative shipments and an infrastructure change}

The route $a_1\to x\to b_1$ costs the same as the direct link $a_1b_1$.
The three units from $a_1$ may be divided in any proportion between these
routes. The five units from $a_2$ still pass through $x$ to $b_2$.
Prices and total welfare remain the same.

Table~\ref{tab:flows} makes this shipment ambiguity explicit. Complete
prices determine which arcs are tight, but quantities and market clearing
are also needed to determine shipment ranges. Even with these data, equally
costly routes can leave individual shipments undetermined.

\begin{table}[t]
\centering
\caption{Arcs with fully observed equilibrium prices.}
\label{tab:flows}
\begin{tabular}{lccc}
\toprule
Arc & Cost & Slack & Flow range\\
\midrule
$a_1\to x$   & 1 & 0 & $[0,3]$\\
$a_2\to x$   & 1 & 0 & $\{5\}$\\
$x\to b_1$   & 1 & 0 & $[0,3]$\\
$x\to b_2$   & 1 & 0 & $\{5\}$\\
$a_1\to b_1$ & 2 & 0 & $[0,3]$\\
$b_1\to b_2$ & 1 & 1 & $\{0\}$\\
$b_2\to b_1$ & 1 & 1 & $\{0\}$\\
\bottomrule
\end{tabular}
\par\smallskip\noindent
\parbox{.94\columnwidth}{\footnotesize The other five reverse arcs have
positive slack and zero flow. Flow ranges use supplies and deliveries
$(3,5)$ and are linked by market clearing.}
\end{table}

Now let the direct-link cost exceed $2$, with all other costs fixed. The route through
$x$ still costs $2$. Equilibrium prices, consumption, and welfare are
unchanged, but all three units from $a_1$ now travel through $x$.
Both directions of the direct link are excluded because its cost exceeds
the absolute price difference.
Infrastructure changes can therefore redirect shipments while leaving
the observed node prices unchanged.

\subsection{Partial observations}

Returning to the original transport costs, suppose only $p_{a_1}=1$ and
$p_{b_1}=3$ are observed. The first observation and the link $a_1x$ require
$p_x\le2$. The second observation and the link $xb_1$ require $p_x\ge2$.
Consequently $p_x=2$. Both arcs on this cheapest route must be tight.
The other unobserved prices have ranges $p_{a_2}\in[1,3]$ and
$p_{b_2}\in[2,3]$.

For example, on $a_2\to x$ the price difference ranges from $-1$ to $1$.
The arc cost is $1$, so slack ranges from $0$ to $2$. Its status is
unresolved. On $b_1\to b_2$ the price difference lies between $-1$ and $0$,
so slack lies between $1$ and $2$ and the arc is excluded. The complete
classification is summarized in Table~\ref{tab:partial}.

\begin{table}[t]
\centering
\caption{Arc information from prices only at $a_1$ and $b_1$.}
\label{tab:partial}
\begin{tabular}{lcl}
\toprule
Arc & Slack interval & Status\\
\midrule
$a_1\to x$   & $[0,0]$ & Necessarily tight\\
$x\to b_1$   & $[0,0]$ & Necessarily tight\\
$a_1\to b_1$ & $[0,0]$ & Necessarily tight\\
$a_2\to x$   & $[0,2]$ & Unresolved\\
$x\to a_2$   & $[0,2]$ & Unresolved\\
$x\to b_2$   & $[0,1]$ & Unresolved\\
$b_2\to b_1$ & $[0,1]$ & Unresolved\\
\bottomrule
\end{tabular}
\par\smallskip\noindent
\parbox{.94\columnwidth}{\footnotesize The other five directed arcs have
positive minimum slack and are excluded. Necessarily tight arcs may carry
zero shipment.}
\end{table}

The fields $(1,1,2,3,3)$ and $(1,3,2,3,2)$ both match the two observations
and satisfy arbitrage. Their tight arcs differ. In the second field,
the tight direction on $a_2x$ is reversed. The known economic primitives
rule out that second field as an equilibrium: the producer at $a_2$ has
a positive price and supplies five units, yet its only outgoing arc
is slack. This separates the information provided by prices and costs
from the additional information provided by demand and production.

Starting from the observations in Table~\ref{tab:partial}, an observation at
$x$ supplies no further restriction, since its price is already determined.
Adding $p_{b_2}=3$ makes $x\to b_2$ necessarily tight and excludes
$b_2\to b_1$, reducing the number of unresolved arcs from four to two.

\subsection{Spatial selection and numerical checks}

For the five-node field, we use planar coordinates
$a_1=(0,1.6)$, $a_2=(0,0)$, $x=(1.8,0.8)$,
$b_1=(3.6,1.6)$, and $b_2=(3.6,0)$, with triangles
$(a_1,a_2,x)$, $(a_1,x,b_1)$, $(a_2,b_2,x)$, and $(x,b_2,b_1)$.
The interpolated price field is $p(X,Y)=1+5X/9$, with constant slope and
energy $8/9$. The two-stage calculation retains
these prices and the economic value $J^*=20$.

A separate three-vertex example shows what spatial selection does when
an equilibrium leaves a junction price undetermined. We place a producer,
consumer, and junction at $(0,0)$, $(1,0)$, and $(0,1)$. Producer capacity
is $1$; consumer mass is $2$ with uniform valuations on $[0,4]$.
The producer--consumer edge costs $1$ and the producer--junction edge
costs $10$. There are no other edges. Equilibrium fixes the producer
price at $1$ and the consumer price at $2$; the junction price $q$ can
lie anywhere in $[-9,11]$ without changing the economic objective $J^*=2$.
On this triangle the spatial energy is
\begin{equation}\label{eq:selection-example}
 \mathcal E_h(q)=\frac{1+(q-1)^2}{4}.
\end{equation}
It is smallest at $q=1$, with value $1/4$. For comparison, the equally
valid junction price $q=1.8$ gives energy $0.41$. The second minimization
chooses the smoother field while preserving equilibrium and welfare.

We solve the welfare and price problems independently
and verify the balance, capacity, demand, and arc conditions. Sequential least squares
programming (SLSQP) is used for price optimization, with a linear refinement of
the binding conditions before spatial selection; HiGHS is used for the arc
bounds. In the five-node baseline, the absolute welfare--price gap is below
$10^{-10}$.
Multiplying both producer capacities by ten gives producer prices of zero
and consumer prices of $2$, with some capacity unused. Raising the direct-link cost
from $2$ to $3$ preserves welfare and removes use of that link.

The price-difference bounds are checked on all $31$ nonempty observation
subsets of the five-node network, using both the baseline and the changed
direct-link cost. For every subset, both slack endpoints on all $12$
directed arcs are compared with values obtained from independently solved
linear programs.
All $1,488$ comparisons agree within $10^{-7}$. Observations $p_{a_1}=1$
and $p_{b_1}=4$ are rejected because their difference exceeds the
shortest-path cost.

A planar interpolation check uses a $3\times3$ vertex grid on the unit
square, split into eight triangles. Boundary prices are sampled from
$p(X,Y)=1+X$, and each mesh edge has transport cost twice its length.
Minimum-energy interpolation recovers the central price $1.5$ and energy
$1/2$, with a zero interior Euler--Lagrange residual to numerical
precision.

\section{Discussion and Future Work}

The model forms part of our ongoing research into the economic importance of
production and transportation nodes. We are developing applications to node
disruption, in which available production capacity or transport connections
are reduced and the equilibrium is recalculated. These applications will
compare changes in delivered prices, consumption, and total economic surplus
across the region. We use the surviving physical network, allowing alternative
routes to carry shipments as prices adjust. We can then assess a node's
importance through the economic consequences of its disruption.

Production and transport functions are treated separately, and demand at a
disconnected location remains part of the economy. If a closure disconnects
the network, each connected component requires a separate equilibrium
calculation. These comparisons concern equilibria before and after disruption;
adjustment over time is a further question.

A further application concerns incomplete observations. The price bounds
developed here allow us to examine how the information available about feasible
shipments changes as prices are observed at additional locations. This provides
a basis for investigating where further observations of prices and quantities
would be most informative.

Empirical applications require estimation of demand and transport costs from
regional data. Observed shipments would provide a check on the inferred arc
exclusions and shipment ranges. Further computational work includes spherical
implementation and convergence under mesh refinement with the economic nodes
held fixed.

\section{Conclusion}

We formulate a variational framework for spatial price determination
with discrete production and demand embedded in continuous geography.
The economic price problem determines equilibrium, and a second variational
problem selects its spatial form through minimum energy. Partial price observations and
known transport costs give attainable bounds on unobserved prices and arc
slack under arbitrage. These bounds can certify the exclusion of directed
arcs from equilibrium shipments even when the complete shipment allocation
remains unknown.

\bibliographystyle{IEEEtran}
\bibliography{submission_draft_v9}
\end{document}